# Predicting Short-Term Uber Demand Using Spatio-Temporal Modeling: A New York City Case Study

Sabiheh Sadat Faghih, Abolfazl Safikhani, Bahman Moghimi, Camille Kamga

sfaghih00@citymail.cuny.edu, as5012@columbia.edu, smoghim000@citymail.cuny.edu, ckamga@utrc2.org

**ABSTRACT:**

*The demand for e-hailing services is growing rapidly, especially in large cities. Uber is the first and popular e-hailing company in the United Stated and New York City. A comparison of the demand for yellow-cabs and Uber in NYC in 2014 and 2015 shows that the demand for Uber has increased. However, this demand may not be distributed uniformly either spatially or temporally. Using spatio-temporal time series models can help us to better understand the demand for e-hailing services and to predict it more accurately. This paper analyzes the prediction performance of one temporal model (vector autoregressive (VAR)) and two spatio-temporal models (Spatial-temporal autoregressive (STAR); least absolute shrinkage and selection operator applied on STAR (LASSO-STAR)) and for different scenarios (based on the number of time and space lags), and applied to both rush hours and non-rush hours periods. The results show the need of considering spatial models for taxi demand.*

## 1. INTRODUCTION

With the rise of intelligent transportation systems, e-hailing services (app-based services), have recently become popular among transportation users. Companies, such as Uber, Lyft, Juno, Gett, or Via, that passengers can request rides from phone application are called e-hailing service companies. The market for these services is growing, attracting more customers, and competing fiercely with other ride-hailing services in big cities in the United State, like New York City (NYC). In New York City, the number of trips by street hailing taxis (yellow cabs) has fallen between 2014 and 2015, while, during the same time period, the demand for e-hailing companies such as Uber has increased significantly as shown in Figure 1.

Drivers in hailing services (either e-hailing or street hailing) have to search for their next passenger, which entails driving an empty taxi around the city. At the same time, in some parts of urban areas, passengers may have to wait for a long time to find a cab. It has been shown that having a better knowledge of demand in the near future can improve the efficiency of the system (Qian et al., 2017). It can help drivers reduce their empty cruising by suggesting locations where they might find passengers (Wang et al., 2017, Safikhani et al., 2017) and help passengers reduce their waiting time. Moreover, decreasing empty taxi travel time and distance can result in less congestion and pollution (Li et al., 2017).

In this research, thanks to the availability Uber data and its high number of daily trips, the behavior of Uber pick-up demand is going to be analyzed. It is intuitive that demand for Uber in each area of the city changes from one time interval (15 min) to the next. In addition, the volatility of Uber demand over time differs from one area of the city to another. Such demand is changing both spatially and temporally. One key question that this paper is trying to answer is how much correlation there is in Uber demand from one area/district to another. A new spatial-temporal modeling approach will be used to answer this question and to predict the demand more accurately. This approach will use time series models to capture the time and spatial variation in Uber demand, determine the correlation among the Uber data, and predict future demand in real time.

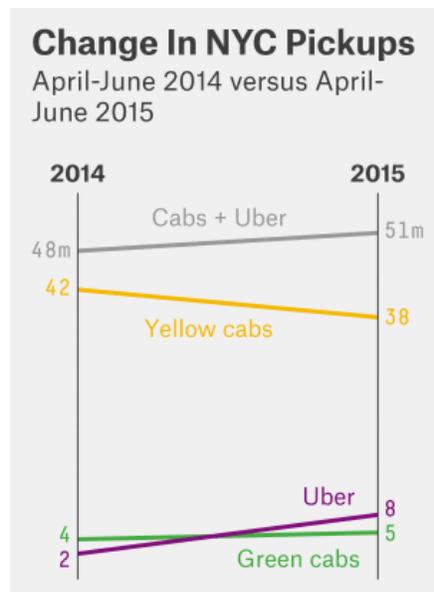

**FIGURE 1 Taxi pick -up changes in Manhattan (Taxi and Limosine Commission).**

In this paper, two days of yellow cab data in Manhattan are analyzed to determine if the demands for such services can be described by spatio-temporal time series models. Among different spatio-temporal models, this paper compares the prediction performance of STAR and LASSO -STAR models for different scenarios, and tries to find parameters that can improve the accuracy of the models. In the context of spatio-temporal time series models, "weight matrices" play an important role in correlating values for one district/zone to those for other areas. The matrix is a representation of how different districts/zones are correlated to one another. Several types of weight matrices are defined in the literature (Qian et al., 2017, Getis 2009, Getis and Altstadt 2010). In this paper, two weight matrices are proposed for use in the prediction models and the one which provides the most accurate prediction is used to analyze the performance of the models applied to rush-hour and non-rush hour demand.

## 2. LITERATURE REVIEW

Initial research about ride-hailing for NYC taxis was undertaken in a study done by Schaller (Qian et al., 2017). To understand the behavior of taxi demand, the study applied a citywide empirical time series regression model in order to better picture the relationship between taxicab revenue per mile and economic activity in the city, taxi supply, taxi fare, and bus fare. With the emergence of GPS technology, Yang and Gonzales (2017) analyzed GPS taxi data in New York City and tried to explain the variation in taxi demand using six explanatory variables including population, education, median income per capita, median age, employment by industry sector, and transit accessibility. Correa et al. (2017) developed demand forecasting models for Uber and taxis to investigate the spatial dependence between trips made by the two services and found a significant correlation between Uber and taxi demand.

Moreira-Matias et al. (2013) developed an approach to predict short-term taxi demand at 30 minute intervals. The approach combined three predictive models, a Time Varying Poisson Model, a Weighted Time Varying Poisson Model, and an Auto-Regressive Integrated Moving Average Model. They found that the results obtained by their combined modeling approach were superior to those obtained by each model separately. A Gaussian Conditional Random Field (GCRF) model was developed by Qian et al. (2017), to predict short-term taxi demand. However, the GCRF model may require t a high dimension of inputs, which may not be available and may cause an over-fitting problem. In light of these problems, the authors proposed a new model using a sparse GCRF and estimating the model's parameters with LASSO for predicting demand. Their model was applied to yellow cab data in Manhattan, NYC in which the pick-up data was aggregated over 15 minute intervals and by zip-code. The proposed model, along with ARIMA and ANN models, were applied to 4 different scenarios to assess its performance. The demand was predicted in 2 and 10 time-steps for peak hours and off-peak hours respectively. It was found that the developed model outperformed the other two algorithms. 15 to 30 min ahead prediction is a realistic opportunity for applying real-time strategy (Koutsopoulos et al., 2017). For temporal analysis, a well-known family of time series called Autoregressive Integrated Moving Average (ARIMA) (Moghimi et al., 2017), as a univariate model, or vector autoregressive (VAR) as a multivariate model, can be beneficial. However, in a dense urban transportation network including many areas that each has its own demand variation, the taxi/Uber demand can be correlated from one to another; hence, the spatial-temporal modeling can better capture the underlying demand.

The idea of using spatial information (considering the spatial correlation among location) in predicting was first appeared in the transportation literature in the study done by Okutani and Stephanedes (1984) for prediction of traffic flow. Later, Kamriankis and Prastacos (2003) took advantage of the using spatial temporal modeling to predict the relative traffic velocity on major roads of Athens, Greece. The approach was called space-time autoregressive integrated moving average (STARIMA). The introduced model was superior to conventional time series models like ARIMA and VARMA. Furthermore, Cheng et al. (2011) developed a STARMA model to determine the dynamic autocorrelations of road network data obtained from the Automatic Number Plate Recognition (ANPR) system in Central London. The results revealed that the proposed model improved the estimation and prediction of traffic in comparison to the traditional time series model, ARIMA. Recently, Duan et al. (2016) developed a STARIMA-based model with time-varying lags to predict short-term traffic flow and the experimental results showed that the developed model had superior accuracy compared with traditional cross-correlation functions and without employing time-varying lags.

In STARMA modeling, the correlation between various areas is represented by a weighting matrix, which is one of the key components of such modeling. The matrix can be fixed or dynamic. Min et al. (2009) presented a Dynamic STARMA model to forecast short-term traffic flow and applied their method to the urban grid of Beijing, China. The matrix's values used in their study were not fixed, rather they changed from time to time depending on the proportion of traffic volume moving from upstream links toward downstream links. The appropriate weighting matrix structure varies based on the nature of each problem. One approach can be to define a ring of dependency by an adjacency order. For instance, a first-order adjacent matrix shows the dependency of those areas to themselves, a second-order adjacent matrix shows the dependency of those area to the target area, and a third-order matrix considers those areas adjacent to the second-order, and indirectly close to the target area. Weighting matrices can include up to fifth or sixth-order dependencies. Kamarianakis et al. (2004) applied first and second order adjacency matrices. Another approach is to consider the distance between areas, and construct different adjacency matrices based on defined threshold values of distance, which seems more practical for transportation networks since the size and shape of zip-codes, census tracts, or transportation analysis districts (TADs) are not uniform.

In the next section, this paper briefly describes the spatio-temporal methodology and the structure of the STAR and LASSO-STAR models. Subsequently, in the section, Implementation and Data Preparation, the implementation of the spatio-temporal models for predicting short-term Uber demand is discussed. This section will also discuss the preparation of Uber pick-up data, the zoning system, and the weight matrices. Afterward, in the results section, the outputs of the spatio-temporal models for Manhattan, New York are presented. The models were applied to a typical day, as well as over two days, to see the impacts on demand during rush and non-rush hours. Finally, the conclusion section presents the main findings and discusses future work.

## 3. METHODOLOGY

In this section, VAR, STAR and LASSO-STAR models will be described briefly. It is referred to (Safikhani et al., 2017) for more details about spatial-temporal modeling. Let's assume showing the number of pick-ups in $i$-th TAD at time $t$ by $Y_i(t)$ and also show the max time lag by $p$ and the number of TADs by $k$.

Suppose $s_1, s_2, \ldots, s_k$ are $k$ fixed locations in $\mathbb{R}^d$ at which the response variable $\{y_t = (y_t(s_1), y_t(s_2), \ldots, y_t(s_k)) \in \mathbb{R}^k\}_{t=1}^T$ has been observed over a period of time with length $T$. Then, $y_t$ is called a VAR model if

$$y_t = \nu + \Phi^{(1)} y_{t-1} + \cdots + \Phi^{(p)} y_{t-p} + u_t, \ t = 1, 2, \ldots, T, \quad (1)$$

where $\nu \in \mathbb{R}^k$ the intercept, $\Phi^{(i)} \in \mathbb{R}^{k*k}$ the i-th lag coefficient matrix, and $\{u_t \in \mathbb{R}^k\}_{t=1}^T$ is a mean zero k-dim white noise with covariance matrix $\Sigma_u$. There are $k(kp+1)$ parameters to estimate, and if $k$ is large compared to T, we may need to reduce the size in our estimation procedure.

The two other models are spatio-temporal models that means they consider the correlation between different districts. The following linear regression

$$Y_i(t) = \sum_{j=1}^P \sum_{l=0}^{\eta_j - 1} \phi_i^{(j,l)} W_i^{(l)} Y(t-j) + \varepsilon_i(t), \quad (2)$$

where $\varepsilon_i(t) = (\varepsilon_1(t), \ldots, \varepsilon_k(t))$ is a k-variate normal variable with mean zero and

$$\mathbb{E}\left(\varepsilon(t)\varepsilon(t+s)'\right) = \begin{cases} \sigma^2 I_k, & s = 0 \\ 0, & \text{otherwise} \end{cases}$$

Also, $W^{(l)}$'s are $k*k$ matrices which govern the $l$-th neighborhood location with $W_i^{(0)} = I_k$. Denote the $i$-th row of $W^{(l)}$ by $W_i^{(l)}$. These matrices are then normalized in such a way that the sum of each row would be 1. Finally, for each $i = 1, 2, \ldots, k$, and $j = 1, 2, \ldots, p$, $\phi_i^{(j, 0: \eta_j - 1)} = \left(\phi_i^{(j,0)}, \phi_i^{(j,1)}, \ldots, \phi_i^{(j, \eta_j - 1)}\right)$ is a vector of coefficients of size $\eta_j$ relating the current observation at location $i$, $Y_i(t)$, to the all weighted observations in $\eta_j$ different neighborhoods $j$ time lags in the past. Without loss of generality, it is assumed that $\eta_1 = \cdots = \eta_p = \eta$. Further, denote $\Phi_i = \left(\phi_i^{(1, 0: \eta - 1)}, \ldots, \phi_i^{(p, 0: \eta - 1)}\right)$. In order to write equation (2) in matrix form, let $Y_i = Y_i(1), \ldots, Y_i(T))$, $\varepsilon_i = (\varepsilon_i(1), \ldots, \varepsilon_i(T))$, and define $Z_i$ to be the $T \times \eta p$ with $Z_i(t, (j-1)\eta + l) = W_i^{(l)} Y(t-j)$ for $t = 1, 2, \ldots, T$, $j = 1, 2, \ldots, p$, and $l = 0, 2, \ldots, \eta - 1$. Now, one can write the data equation for $i$-th time series component as follows:

$$Y_i = Z_i \Phi_i + \varepsilon_i \quad (3)$$

This model reduces the number of parameters from $k^2 * p$ in the VAR model to $k * \eta * p$, assuming $\eta \ll k$. Least squares estimation can be implemented for parameter estimation, i.e. for $i = 1, 2, \ldots, k$,

$$\widehat{\Phi}_i = argmin_{\Phi_i} \frac{1}{2} \|Y_i - Z_i \Phi_i\|_2^2, \quad (4)$$

with $\|.\|_2$ being the Euclidean norm. However, for the cases when $T$ is small compared to $k$, it might be beneficial to still reduce the number of parameters in the model with the goal of improving forecast performance. For that, a penalty function $\Omega(\Phi)$ will be added to equation (4) with the purpose of setting some of the small parameters to zero to increase forecast efficiency. More specifically,

$$\widehat{\Phi}_i = argmin_{\Phi_i} \frac{1}{2} \|Y_i - Z_i \Phi_i\|_2^2 + \lambda \, \Omega(\Phi_i), \tag{5}$$

Where $\lambda$ is the tuning parameter to be selected by cross validation techniques. The penalty function chosen in this article is the well-known LASSO penalty (1996), which is a simple element-wise $L_1$ penalty on all the components of $\Phi_i$, i.e. for $i = 1, 2, \ldots, k$, as can be seen herein

$$\Omega(\Phi_i) = \sum_{j=1}^{p} \sum_{l=0}^{\eta-1} \left| \phi_i^{(j,l)} \right| \tag{6}$$

## 4. IMPLEMENTATION AND DATA PREPARATION

To implement the proposed models, the time interval points are divided into three parts $0 < T_1 < T_2 < T$. For fixed values of $\lambda$, the optimization problem (4) is solved on the interval $[0, T_1]$. Then, the mean squared prediction error (MSPE) for predicting one step ahead is calculated over all $k$ time series components on the second portion of time points which is the time interval $[T_1 + 1, T_2]$. Subsequently, the tuning parameter $\lambda$ which minimizes this MSPE will be selected, and the model performance then can be quantified by the MSPE on the last part of the data, which is on the time interval $[T_2 + 1, T]$. The formula for MSPE is shown in equation (7).

$$MSPE = \frac{1}{k(T_2 - T_1)} \sum_{i=1}^{k} \sum_{t=T_1+1}^{T_2} (Y_i(t) - P_{T_1} Y_i(t))^2, \tag{7}$$

Where $P_{T_1} Y_i(t)$ the best linear predictor of is $Y_i(t)$ based on the first $T_1$ observations

Data preparation is an important step before implementing the models. To predict the number of pick-ups in one district, the STAR and LASSO-STAR models need the history of pick-ups in each district, as well as a weight matrix as inputs. The pick-up data and weighting matrices are discussed next.

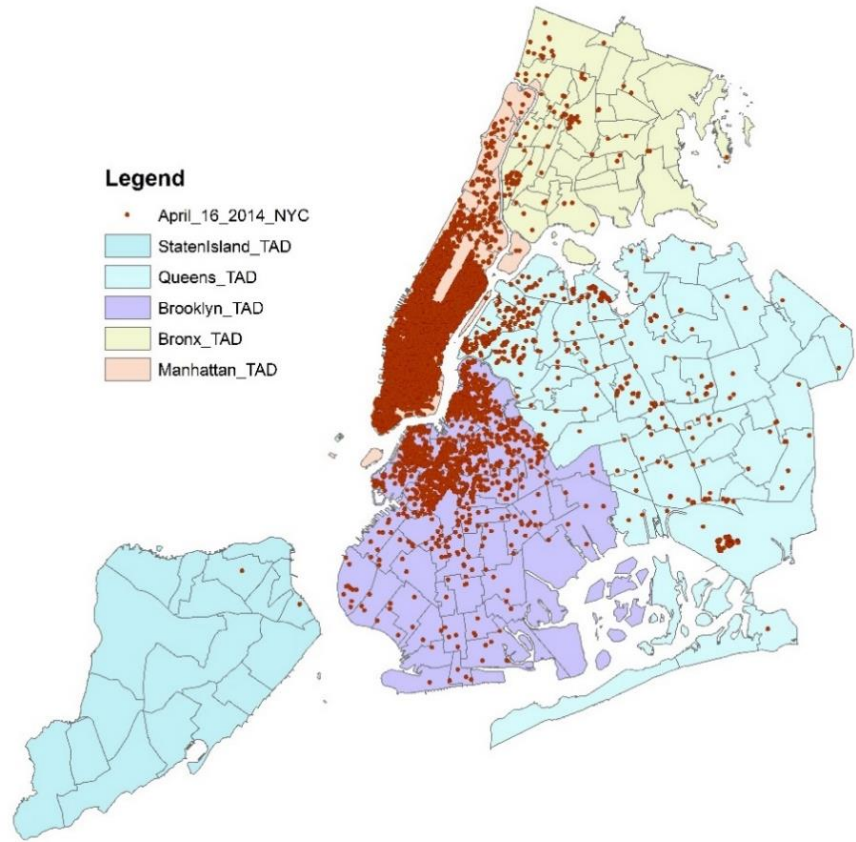

**FIGURE 2 Uber pick-ups in New York City.**

**Uber Pick-up Data**

To study the prediction performance of the models, the pick-up data for a typical day is chosen. Our focus in this paper is to describe how these models can be used for Uber demand prediction. Uber data contains information about the coordination and time of the pick-ups and drop-offs of each trip through a day. Based on available data, we looked at the historical Uber data of Uber from April 2014 through September 2014. The pattern of number of pick-ups stays the same for weekdays, specifically based on their autocorrelation. To test these models, a typical day is picked, however the procedure can be extended for other days. A typical day is usually considered to be Tuesday, Wednesday or Thursday when the schools are open and the weather is not extreme such as during the month of April (Barann, Beverungen, & Müller, 2017, Yazici, Kamga, & Singhal, 2013), September or October (Qian et al., 2017). We selected the 16th and 17th of April 2014 as typical days for this study. Figure 2 shows the pick-up points of the Uber trips on April 16th. The pick-up points were aggregated both spatially and temporally: based on their longitude and latitude, the pick-ups were assigned to Manhattan Traffic Analysis Districts (TAD) and then aggregated to 15-min intervals. The outcome is a 27x96 matrix of the number of pick-ups for each day, whose indices represent the TAD and the time interval.

**Zoning System**

Previously, zip-code-based aggregation was used in a study done by Qian et al. (2017). In this paper, aggregation of Uber pick-up data is based on Manhattan's TADs. The reason is that Manhattan's zip-codes vary in size from very large areas to areas as small as a single building. Figure 3 displays Manhattan's 27 TADs and the centroid of each district.

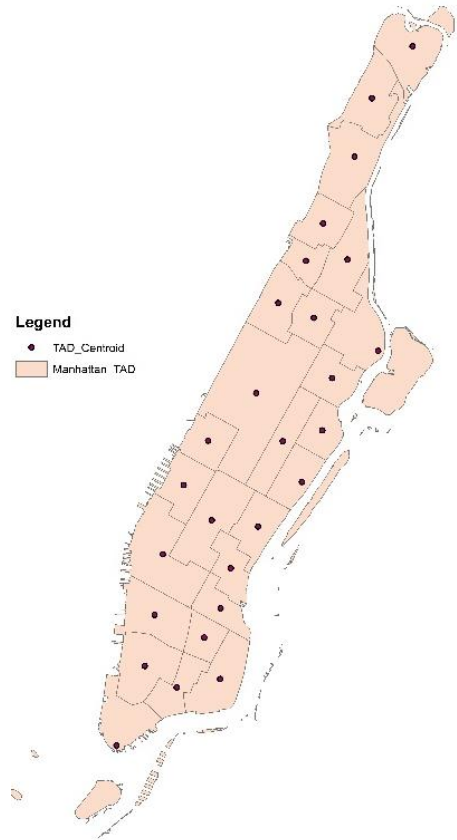

**FIGURE 3 Manhattan's zoning system based on TAD.**

**Weight Matrices**

As mentioned in the Methodology Section, a weight matrix "*reflects a hierarchical order of spatial neighbors*" (Pfeifer and Deutrch, 1980) and, as such, weight matrices are an essential input for spatial models. The collection square matrices governing all the neighborhood lags form the weighting matrix. A detailed discussion of the structure of these matrices can be found in (Pfeifer and Deutrch, 1980). We can assume and assess that districts which are closer together have higher correlation to each other as compared to districts that are farther apart. Two methods are used to order the districts and produce weighting matrices for the TAD zoning system. The two methods are as follows:

1- Based on the distance between centroids: The geometric center of each district is calculated and then the other districts are categorized based on their Euclidean distance from this district. The first order matrix only consists of each district, so the distance is zero. For the second to sixth order matrices, an increasing number of surrounding districts are taken into account. These six square matrixes are combined together and form a weight matrix.

2- Based on the number of neighbors between districts: In this method, we visually determined how many districts are located between two districts. Similar to the previous method, each district is considered as the only district in the first order matrix, so the distance is zero.

## 5. RESULTS

**Part 1 Results: Analysis of One Day**

Considering April 16$^{th}$, 2014, there are 96 points available for each district. As explained in the implementation section, 2/3 ($T_2$) of these points are used for fitting, tuning, and estimating the parameters to predict the last 1/3 of data points. One temporal model (VAR) and two spatio-temporal models (STAR and LASSO-STAR) are run with two different weighting matrices, different time lags ($p = 1, 2, 3, 4$), and various spatial lags ($\eta = 1, 2, 3, 4, 5, 6$). The performance measurements in terms of MSPE for the STAR and LASSO-STAR models are displayed in Table 1.

At first glance, a huge difference between the temporal and spatio-temporal models is observed. Although the VAR model uses $k^2 \times p$ parameters in its estimation, it did not provide a better performance compared to the STAR and LASSO-STAR models that use $k \times \eta \times p$ parameters. This comparison highlights the importance of using spatio-temporal models for predicting taxi demand, which has been recognized by other scholars (Qian et al., 2017, Saadi et al., 2017, Davis, Raina, & Jagannathan 2016).

It can also be noticed from the MSPE results that, in almost all cases, the LASSO-STAR prediction model performs better than the STAR model. The STAR model outperforms the LASSO-STAR model in only two cases both of which are when $\eta = 1$, which means that the spatial effects of other districts are neglected, as $\eta = 1$ refers to the first-order matrix in which no neighbor districts are considered. In these cases, the effect of penalization is negligible due to the low-dimensionality of the model. However, the impact of penalty function is noticeable on cases with high time lags ($p = 4$).

Of the 48 combinations of spatial and time lags and weighting matrix types, the LASSO-STAR model performs the best as indicated by the lowest MSPE value when $p = 1, \eta = 6$ and the weighting matrix based on the number of neighboring districts is used. The LASSO-STAR model is successful in controlling the number of coefficients, so it can easily consider high levels of spatial lags ($\eta = 6$) without worsening the accuracy of the model. On the other hand, the STAR model's performance decreases as the number of spatial lags increases for both types of weighting matrices. Thus, the model's best performance occurs when the spatial effects of other districts are neglected ($\eta = 1$) and data from one more time lag is considered ($p = 2$).

Considering Table 1, it is clear that the performance of the models also depends on the weighting matrices. Between two introduced weighting matrices, $W = W_2$ could better capture the spatial structure, having higher accuracy. It is worth noting that the performance of proposed models using $W_1$ weighting matrix is reasonably well specially comparing to VAR model. $W_1$ was produced simply based on the distances between center of the districts, while for $W_2$ for each district, it is visually specified the districts at its *n-th* spatial lag. That can be part of the reason why $W_2$ is associated with more accurate prediction.

| Space Lag | Model | Using W1 as Weighting Matrix, (Based on the centroid distances) | | | | Using W2 as Weighting Matrix, (Based on Neighboring Level) | | | |
|---|---|---|---|---|---|---|---|---|---|
| | | P=1 | P=2 | P=3 | P=4 | P=1 | P=2 | P=3 | P=4 |
| $\eta = 6$ | STAR | 1.0251 | 1.4952 | 3.1445 | 9.7788 | 1.1806 | 1.8273 | 2.7657 | 10.4838 |
| | LASSO-STAR | 0.9142 | 0.9794 | 0.9693 | 1.0804 | 0.9028 | 1.0408 | 1.0478 | 1.0854 |
| $\eta = 5$ | STAR | 1.0191 | 1.3084 | 2.0834 | 4.9545 | 1.1735 | 1.7474 | 2.4568 | 3.8210 |
| | LASSO-STAR | 0.9221 | 1.0393 | 1.0558 | 1.1266 | 0.9776 | 0.9844 | 1.1147 | 1.1462 |
| $\eta = 4$ | STAR | 1.0020 | 1.1944 | 1.7242 | 2.5057 | 1.0719 | 1.2908 | 1.6918 | 2.8634 |
| | LASSO-STAR | 0.9077 | 0.9064 | 0.9464 | 1.0945 | 0.9402 | 0.9818 | 0.9766 | 1.1420 |
| $\eta = 3$ | STAR | 0.9714 | 1.1660 | 1.5077 | 2.1659 | 0.9824 | 1.1512 | 1.4610 | 1.9709 |
| | LASSO-STAR | 0.9182 | 0.9379 | 0.9573 | 1.0822 | 0.9457 | 0.9598 | 0.9558 | 1.0487 |
| $\eta = 2$ | STAR | 0.9525 | 0.9924 | 1.1274 | 1.4069 | 0.9486 | 0.9985 | 1.1469 | 1.3762 |
| | LASSO-STAR | 0.9355 | 0.9182 | 0.9353 | 0.9594 | 0.9295 | 0.9197 | 0.9411 | 0.9806 |
| $\eta = 1$ | STAR | 0.9664 | 0.9124 | 0.9515 | 1.0232 | 0.9664 | 0.9124 | 0.9515 | 1.0232 |
| | LASSO-STAR | 0.9290 | 0.9381 | 0.9575 | 0.9396 | 0.9290 | 0.9381 | 0.9575 | 0.9396 |
| | VAR Model | 8.5355 | 1.7410 | 1.3622 | 1.2354 | 8.5355 | 1.7410 | 1.3622 | 1.2354 |

**TABLE 1 Performance Measurements (MSPE) for LASSO-STAR, STAR and VAR Model with Different η and P**

## Part 2 Results: Analysis of Rush and Non-Rush Hours

To better understand the behavior of Uber demand during different times of the day, the models' performance is analyzed during rush hours and non-rush hours. These two time intervals are selected from the next day (April 17[th], 2014, Thursday), since time series models need a reasonable history to calculate more accurate parameters.

The New York City Metropolitan Transportation Authority (MTA) considers the morning rush hour to be the three hours between 6:30 and 9:30 AM and the afternoon rush hour to be from 3:30 to 6:30 PM. Uber Data for April 17[th], was aggregated as described above and combined with the data for April 16[th]. To develop models for rush hour and non-rush hour demand, the following two time intervals are defined:

1- Morning rush hour: 6:30am ~ 9:30 am
2- Midday non-rush hour: 9:30am~12:30 pm

For the morning rush hour, the time series is constructed from 12:00 AM April 16thto 9:30 AM April 17th. Since we are interested in estimating the demand during rush hour, the value of T is the time lag associated with 9:30 AM and $T_2$ is associated with the 6:30 AM time lag. $T_1$ is easily set as one half of $T_2$ (means the time intervals from 12:00 AM to 3:15 AM). The same logic is applied for the second time interval for the non-rush hour: T as 12:30 PM and $T_2$ as 9:30 AM.

It was shown in the previous analysis that considering time lags as large as 3 or 4 increases the prediction error, also the best results happened with the $W_2$ (neighboring level matrix). So, in this analysis, the models are tested for time lags $p = 1, 2$ and $3$ with $W_2$ as the weighting matrix. Tables 2 and Table 3 display the performance measurements for the LASSO-STAR and STAR models for the rush hour and non-rush hour respectively. (Time lags: $p = 1, 2, 3$ spatial lags $\eta = 1, 2, 3, 4, 5, 6$; $W = W_2$).

| | Space Lag | Model | P=1 | P=2 | P=3 |
|---|---|---|---|---|---|
| Morning Rush Hour (6:30 am~ 9:30 am) | $\eta = 6$ | STAR | 0.9555 | 1.0626 | 1.2291 |
| | | LASSO-STAR | 0.9383 | 0.9585 | 0.9570 |
| | $\eta = 5$ | STAR | 0.9717 | 1.0775 | 1.2205 |
| | | LASSO-STAR | 0.9625 | 0.9803 | 0.9825 |
| | $\eta = 4$ | STAR | 0.9572 | 1.0245 | 1.1200 |
| | | LASSO-STAR | 0.9467 | 0.9674 | 0.9803 |
| | $\eta = 3$ | STAR | 0.9665 | 1.0178 | 1.0724 |
| | | LASSO-STAR | 0.9695 | 0.9895 | 1.0063 |
| | $\eta = 2$ | STAR | 0.9596 | 0.9667 | 0.9902 |
| | | LASSO-STAR | 0.9735 | 0.9750 | 0.9844 |
| | $\eta = 1$ | STAR | 1.0298 | 0.9875 | 0.9996 |
| | | LASSO-STAR | 1.0312 | 0.9877 | 0.9993 |
| | | VAR Model | 5.3801 | 3.4909 | 1.7738 |

**TABLE 2 Performance Measurements (MSPE) of the Models in Predicting the Demand during Rush Hour**

Similar to what was found in the Part 1 analysis, the performance measurement in STAR and LASSO-STAR are far better than VAR, and also in almost all cases, the LASSO-STAR model provides a better prediction than the STAR. During the morning rush hour, the LASSO-STAR model with $p = 1, \eta = 6$ has the lowest MSPE, while during midday, the LASSO-STAR model with $p = 2, \eta = 2$ outperforms the other cases. During non-rush hours, the demand is more static, which indicates the demand of current time lag depends on higher previous time lags rather than higher neighborhood lags. In other words, during this time, the demand values of each district show almost no correlation with farther districts, but instead the demand is correlated with its own previous values. To summarize, based on the results above, during the non-rush hour, districts (TADs) tend to behave as if they are isolated with demand that is little affected by the demands in other districts, while, during the rush hour, the districts' demands are affected even by their far away neighbors.

| | Space Lag | Model | P=1 | P=2 | P=3 |
|---|---|---|---|---|---|
| Mid-day Non-Rush Hour (9:30 am~ 12:30 pm) | $\eta = 6$ | STAR | 0.3921 | 0.4360 | 0.4974 |
| | | LASSO-STAR | 0.3815 | 0.3681 | 0.3804 |
| | $\eta = 5$ | STAR | 0.4011 | 0.4427 | 0.4782 |
| | | LASSO-STAR | 0.3909 | 0.3774 | 0.3764 |
| | $\eta = 4$ | STAR | 0.4092 | 0.4362 | 0.4579 |
| | | LASSO-STAR | 0.3990 | 0.3900 | 0.3877 |
| | $\eta = 3$ | STAR | 0.3884 | 0.4061 | 0.4363 |
| | | LASSO-STAR | 0.3853 | 0.3721 | 0.3707 |
| | $\eta = 2$ | STAR | 0.3789 | 0.3662 | 0.3758 |
| | | LASSO-STAR | 0.3764 | 0.3567 | 0.3577 |
| | $\eta = 1$ | STAR | 0.4033 | 0.3786 | 0.3897 |
| | | LASSO-STAR | 0.4010 | 0.3753 | 0.3829 |
| | | VAR Model | 2.8876 | 1.0762 | 1.0762 |

**TABLE 3 Performance Measurements (MSPE) of the Models in Predicting the Demand during Non-rush Hour**

For different time and spatial lags, it is noticeable that the MSPE values for the rush hour are much larger than the non-rush hour MSPE values for the corresponding time and spatial lags. It is also worth noting that the value of MSPE in Table 1 for each case lies between the MSPE values for the rush hour and non-rush hour. During the rush hour, the variability of demand is larger. For example, a prediction with a 10% error, will add 1 unit to the squared error if the actual demand is 10, while, with an actual demand of 50, the squared error would increase by 25 units. That is why the models show a better performance in the non-rush hours with values of MSPE decreasing to around 0.3.

## 6. CONCLUSION

This paper introduces a new modeling approach for capturing e-hailing service demand, specifically Uber demand, in Manhattan, New York City. Uber pick-up data is aggregated to the Manhattan TAD level and to 15-min time intervals. This aggregation enables a new spatio-temporal modeling approach to be applied to gain an understanding of demand both spatially and temporally. Two spatio-temporal models, LASSO-STAR and STAR, were developed using Uber pick-up data over a typical day and the performance of the models was measured by MSPE. The MSPE results revealed that it is highly recommended to use the LASSO-STAR model rather than the STAR model. Meanwhile, the knowledge of demand information in surrounding areas can improve the prediction accuracy of the developed spatio-temporal time series models. It is also found that, in spatio-temporal modeling, the type of weighting matrix used can also improve the models' performance. As a continuation of this research, the impact of Uber on yellow taxis will be studied using a change-point detection technique. Moreover, further studies will include additional travel demand-related information such as subway and bus ridership, bicycle demand, weather, etc., as exogenous variables to the time series models.

## ACKNOWLEDGMENTS

The author would like to thank Dr. Ellen Thorson for her valuable comments.

# REFERENCES

1. Barann, B., Beverungen, D., & Müller, O. (2017). An open-data approach for quantifying the potential of taxi ridesharing. Decision Support Systems.

2. Cheng T., Wang J., Harworth J. Heydecker B.G., and Chow A.H.F. (2011). Modeling Dynamic Space-Time Autocorrelation of Urban Tranport Network. GeoComputation, Session 5A: Network Complexity.

3. Correa D., Xie K., Ozbay K. (2017). Exploring the Taxi and Uber Demands in New York City: An Empirical Analysis and Spatial Modeling. Transportation Research Board's 96th, Annual Meeting, Washington, D.C.

4. Davis, N., Raina, G., & Jagannathan, K. (2016, November). A multi-level clustering approach for forecasting taxi travel demand. In Intelligent Transportation Systems (ITSC), 2016 IEEE 19th International Conference on (pp. 223-228). IEEE.

5. Duan P., Mao G., Zhang C., and Wang S., (2016). STARIMA-based Traffic Prediction with Time-varying Lags. IEEE 19th International Conference on Intelligent Transportation System (ITSC), Rio, Brazil.

6. Getis, A. (2009). Spatial weights matrices. Geographical Analysis, 41(4), 404-410.

7. Getis, A., & Aldstadt, J. (2010). Constructing the spatial weights matrix using a local statistic. In Perspectives on spatial data analysis (pp. 147-163). Springer Berlin Heidelberg.

8. Kamarianakis Y. and Prastacos P. (2003). Forecasting traffic flow conditions in an urban network: comparison of multivariate and univariate approaches. Transportation Research Record: Journal of the Transportation Research Board, (1857):74–84.

9. Kamarianakis Y., Prastacos P., and Kotzinos D. (2004). Bivariate traffic relations: A space-time modeling approach. AGILE proceedings, pages 465–474.

10. Koutsopoulos, H. N., Noursalehi, P., Zhu, Y., & Wilson, N. H. (2017, June). Automated data in transit: Recent developments and applications. In Models and Technologies for Intelligent Transportation Systems (MT-ITS), 2017 5th IEEE International Conference on (pp. 604-609). IEEE.

11. Li, Y., Lu, J., Zhang, L., & Zhao, Y. (2017). Taxi booking mobile app order demand prediction based on short-term traffic forecasting. Transportation Research Record: Journal of the Transportation Research Board, (2634), 57-68.

12. Min X., Hu J., Chen Q., Zhang T., and Zhang Y. (2009). Short-Term Traffic Flow Forecasting of Urban Network Based on Dynamic STARIMA Model. Proceedings of the 12th International IEEE Conference on Intelligent Transportation Systems, St. Louis, MO, USA, October 3-7.

13. Moghimi, B., Safikhani, A., Kamga, C., & Hao, W. (2017). Cycle-Length Prediction in Actuated Traffic-Signal Control Using ARIMA Model. Journal of Computing in Civil Engineering, 32(2), 04017083.

14. Moreira-Matias L., Gama J., Ferreira M., Mendes-Moreira J., and Damas, L. (2013). Predicting Taxi-Passenger Demand using Streaming Data. IEEE Transactions on Intelligent Transportation Systems, Volume 14, Issue: 3, DOI: 10.1109/TITS.2013.2262376

15. Okutani I. and Stephanedes Y.J. (1984). Dynamic prediction of traffic volume through kalman filtering theory. Transportation Research Part B: Methodological, 18(1):1–11.

16. Pfeifer, P. E., & Deutrch, S. J. (1980). A three-stage iterative procedure for space-time modeling phillip. Technometrics, 22(1), 35-47.

17. Qian X., Ukkusuri S.V., and Yang C., Yan F. (2017). A Model for Short-Term Taxi Demand Forecasting Accounting for Spatio-Temporal Correlations. Transportation Research Board Annual 2017, Washington D.C.

18. Saadi, I., Wong, M., Farooq, B., Teller, J., & Cools, M. (2017). An investigation into machine learning approaches for forecasting spatio-temporal demand in ride-hailing service. arXiv preprint arXiv:1703.02433.

19. Safikhani, A., Kamga, C., Mudigonda, S., Faghih, S. S., & Moghimi, B. (2017). Spatio-temporal modeling of yellow taxi demands in New York City using Generalized STAR models. arXiv preprint arXiv:1711.10090.

20. Schaller B. (1999). Elasticities for taxicab fares and service availability. Transportation, 26:283–297, DOI: 10.1023/A:1005185421575

21. Taxi and Limosine Commission, https://fivethirtyeight.com/features/uber-is-taking-millions-of-manhattan-rides-away-from-taxis/. Accessed July15th, 2017.

22. Tibshirani, R., (1996) Regression shrinkage and selection via the lasso. Journal of the Royal Statistical Society. Series B (Methodological), pp.267-288.

23. Wang, D., Cao, W., Li, J., & Ye, J. (2017). DeepSD: Supply-Demand Prediction for Online Car-Hailing Services Using Deep Neural Networks. Data Engineering (ICDE), 2017 IEEE 33rd International Conference on (pp. 243-254).

24. Yang C., Gonzales E. (2017). Modeling Taxi Demand and Supply in New York City Using Large-Scale Taxi GPS Data. Seeing Cities Through Big Data - Research, Methods and Applications in Urban Informatics, pp 405-425, DOI10.1007/978-3-319-40902-3_22.

25. Yazici, M. A., Kamga, C., & Singhal, A. (2013, October). A big data driven model for taxi drivers' airport pick-up decisions in new york city. In Big Data, 2013 IEEE International Conference on (pp. 37-44). IEEE.